\documentclass[12pt]{article}

\newif\iffigs\figstrue

\usepackage{epsfig,latexsym}
\usepackage{amsmath}
\usepackage{color}
\usepackage{graphicx}
\usepackage{verbatim}
\usepackage{mathrsfs}
\usepackage{amssymb}

\DeclareMathAlphabet{\mathpzc}{OT1}{pzc}{m}{it}

 \csname
@addtoreset\endcsname{equation}{section}

\def\gz0{\gamma^{0}}

\def\beq{\begin{equation}}
\newcommand{\eeq}[1]{\label{#1}\end{equation}}
\def\bea{\begin{eqnarray}}
\newcommand{\eea}[1]{\label{#1}\end{eqnarray}}
\def\ba{\begin{array}}
\def\ea{\end{array}}
\def\bec{\begin{center}}
\def\ec{\end{center}}
\def\ba{\begin{align}}
\def\ena{\end{align}}

\def\12{\frac{1}{2}}

\usepackage{slashed}

\usepackage{float}

\usepackage{booktabs}
\usepackage{caption}

\usepackage[mathscr]{euscript}

\usepackage{color}
\usepackage{graphicx}
\usepackage{epsf}
\usepackage{graphicx,epsfig}
\usepackage{amsmath}

\usepackage{multirow}

\usepackage{amsmath, amsthm, amssymb}

\usepackage{epsfig}
\usepackage{cite}
\usepackage{color,colordvi}

\newcounter{hran}

\makeatletter
\renewcommand\section{\@startsection {section}{1}{\z@}%
                               {-3.5ex \@plus -1ex \@minus -.2ex}%
                               {2.3ex \@plus.2ex}%
                               {\normalfont\large\bfseries}}
\makeatother

\vspace{0.5cm}

\newcommand{\bi}{\begin{itemize}}
\newcommand{\ei}{\end{itemize}}

\begin{document}
\thispagestyle{empty}
\begin{flushright}
CERN-TH-2017-005
\end{flushright}

\vspace{15pt}

\begin{center}


{\Large\sc Supergravity at 40: Reflections and Perspectives}\\


\vspace{35pt}
{\sc S.~Ferrara${}^{\; a,b,c}$ and A.~Sagnotti${}^{\; d}$}\\[15pt]

{${}^a$\sl\small Theoretical Physics Department, CERN\\
CH - 1211 Geneva 23, SWITZERLAND \\ }
\vspace{6pt}

{${}^b$\sl\small INFN - Laboratori Nazionali di Frascati \\
Via Enrico Fermi 40, I-00044 Frascati, ITALY}\vspace{6pt}

{${}^c$\sl\small Department of Physics and Astronomy \\ Mani L.~Bhaumik Institute for Theoretical Physics, \\ U.C.L.A., Los Angeles CA 90095-1547,
USA}\vspace{6pt}

{${}^d$\sl\small Scuola Normale Superiore and INFN,\\
Piazza dei Cavalieri 7\
I-56126 Pisa, Italy}

\vspace{45pt}

{\sc\large Abstract} \end{center}
\noindent The fortieth anniversary of the original construction of Supergravity provides an opportunity to combine some reminiscences of its early days with an assessment of its impact on the quest for a quantum theory of gravity.
\baselineskip=14pt

\vskip 3 truecm
\begin{center}
{\sl Dedicated to John~H.~Schwarz on the occasion of his 75-th birthday}
\end{center}

\vskip 2.5 truecm
{\sl
\noindent \small Based in part on the talk delivered by S.~F. at the ``Infeld Colloquium and Discrete'', in Warsaw, on December 1 2016, and on a joint CERN Courier article. }
\vfill
\noindent
\baselineskip=20pt
\setcounter{page}{1}

\newpage

\tableofcontents
\newpage

\section{Introduction}
The year 2016 marked the fortieth anniversary of the discovery of Supergravity (SGR) \cite{sugra}, an extension of Einstein's General Relativity \cite{gr} (GR) where Supersymmetry, promoted to a gauge symmetry, accompanies general coordinate transformations. Supersymmetry, whose first realization in four--dimensional Quantum Field Theory was introduced by Julius Wess and Bruno Zumino in \cite{susy,susy_developments}, extends the very notion of spacetime, adjoining to the Poincar\'e group of translations and Lorentz rotations new symmetries that change the Statistics (Bose-Einstein vs Fermi-Dirac) of particles and fields. A peculiar mathematical structure, called ``super-algebra'', achieves this goal while circumventing classic no-go theorems that constrain attempts to unify space-time symmetries (connected to mass and spin) with internal ones (connected to charges of various types) \cite{cm}~\footnote{Supersymmetry was inspired by ``dual resonance models'', an early version of String Theory pioneered by Gabriele Veneziano \cite{ven} and extended by Andr\'e Neveu, Pierre Ramond and John Schwarz \cite{nsr}. Earlier work done in France by Jean-Loup Gervais and Benji Sakita \cite{gs}, and in the Soviet Union by Yuri Golfand and Evgeny Likhtman \cite{gl} and by Dmitry Volkov and Vladimir Akulov \cite{va}, had anticipated some salient features.}. Supergravity implies the existence of a new type of elementary quantum of gravitational origin, a spin 3/2 particle called gravitino.

An exact Supersymmetry would require the existence of super-partners in the Standard Model of Electroweak and Strong interactions and for the gravitational field, but it would also imply mass degeneracies between the known particles and their super-partners. This option has been ruled out, over the years, by several experiments, and therefore Supersymmetry can be at best broken, with super-partner masses that seem to lie beyond the TeV energy region currently explored at the CERN LHC. In Supergravity one would expect that the breaking be spontaneous, as in the Brout-Englert-Higgs (BEH) mechanism of the standard Model \cite{beh}, which was remarkably confirmed by the 2012 discovery of a Higgs particle \cite{lhc}.

Supersymmetry would have dramatic consequences. It would affect the subatomic world, via supersymmetric extensions of the Standard Model (MSSM \cite{mssm} and alike), but also large-scale phenomena and the cosmological evolution of our Universe \cite{cosmology}. Supergravity has the potential to provide important clues for dark matter, dark energy and inflation \cite{inflation}, and for the links between the corresponding breaking scale and the one that ought to have superseded it and presumably still characterizes the present epoch. The recent discovery of gravitational waves from black-hole (BH) mergers is a stunning confirmation of GR \cite{gravitywaves}, the gauge theory of the gravitational field, and one can dream of future revelations of its spin-two quantum, the graviton. The gravitino ought to acquire mass via a supersymmetric version of the BEH mechanism, whose details would also control other mass splittings of crucial importance for super-particle searches (squarks, gluinos, sleptons…).

Let us now describe some key steps in the development of Supergravity, with an eye to achievements and difficulties of this endeavor and to its impact on different fields.

\section{The Early Times}

The first instance of Supergravity was built in the spring of 1976 by Daniel Freedman, Peter van Nieuwenhuizen and one of us (S.~F.) \cite{ffvn}, in a collaboration that had started in the fall of 1975 in Paris, at \'Ecole Normale Sup\'erieure. The construction relied on the vierbein formulation of General Relativity and on the Noether method, an iterative procedure that would result in the non-linear Yang-Mills or Einstein-Hilbert action principles if applied to gauge theories or gravity. Inconsistencies, if present, would have led to obstructions that no further modifications could have overcome. Shortly thereafter, Stanley Deser and Bruno Zumino recovered the result in a simpler and elegant way \cite{dz}, extending the first-order (Palatini) formalism of General Relativity~\footnote{Alternative approaches were soon developed, including Supergravity as the gauge theory of the anti-deSitter group \cite{mmans}, Supergravity on a group manifold \cite{torino} and Supergravity from broken superconformal symmetry \cite{broken_superconformal}.}. In their work the authors of~\cite{dz} focussed on supergravity as a way to bypass inconsistencies of the Velo--Zwanziger type \cite{vz}, which generally affect theories with higher spin fields~(for recent reviews see~\cite{higher_spins}).

These original developments are well captured by eq.~(\ref{eq1}) below, where we display the Lagrangian of $N=1$ Supergravity in four dimensions in the ``1.5 order'' formalism~\cite{sugra}, the torsion equation for the gravitino field $\psi_\mu$ and the supersymmetry transformations for the vierbein field $e^\mu_a$ and $\psi_\mu$ (in ``mostly plus'' signature, as in other examples below)~\footnote{For convenience, we use in all equations the conventions set out by Daniel Freedman and Antoine Van Proeyen in \cite{sugra}. Notice that their definition of $\overline{\psi}_\mu$, say, differs from the usual one, since it includes an imaginary factor $i$.}:
$$ {\cal S} \ = \ \frac{1}{2\,k^2}
\int d^4 x \ e\, \left[ \ e^\mu_a\, e^\nu_b \, {R_{\mu\nu}}^{ab}(\omega) \ - \ \overline{\psi}_\mu\, \gamma^{\mu\nu\rho}\, D_\nu(\omega)\,\psi_\rho \right] \ , $$
\begin{equation} \frac{\delta\, {\cal S}}{\delta \, \omega} \ = \ 0 \ \longrightarrow \
D_\mu\, e_\nu^a \ - \ D_\nu\, e_\mu^a \ = \ \frac{1}{2} \ \overline{\psi}_\mu \, \gamma^a\, \psi_\nu \ , \label{eq1} \end{equation}
$$ \delta\, e_\mu^a \ = \ \frac{1}{2}\ \overline{\epsilon} \, \gamma^a \, \psi_\mu \ , \qquad \quad \delta\, \psi_\mu \ = \ \ D_\mu\, \epsilon \ .$$

Further simplifications of the procedure emerged once its full significance was better appreciated. A mixed formalism was eventually adopted, where it became far simpler to track unwanted terms, and during a Summer Institute held at \'Ecole Normale in August 1976 the Noether procedure led to the first matter couplings \cite{firstmattersugra}, which opened the way to a host of more complicated examples. Moreover, the ``spinning string'' \cite{nsr}, or String Theory as it is now called \cite{strings}, was connected to space-time Supersymmetry via a Gliozzi-Scherk-Olive (or GSO) projection \cite{gso}.  A first extended version of four--dimensional Supergravity, involving two gravitinos, came to light shortly thereafter \cite{n2sugra}.

The low-energy spectra that emerged from the GSO projection pointed to yet unknown ten-dimensional versions of Supergravity, including the counterparts of several gravitinos \cite{su4_sugra}, and to a four-dimensional supersymmetric Yang-Mills theory (SYM) invariant under four distinct supersymmetries \cite{gso,n4ym}.

When S.~F. visited Caltech in the Fall of 1976, he became aware that Murray Gell-Mann had worked out many consequences of Supersymmetry, including upper bounds on the number of gravitinos and on the gauge symmetries allowed, in principle, in all instances of ``pure'' Supergravity, where all particles would be connected to the graviton \cite{grs} (see Table \ref{tab:tab1}). Gell-Mann had realized, in particular, that the largest theory would include eight gravitinos, and would allow for a maximal gauge group, $SO(8)$, which would not suffice to accommodate the $SU(3) \times SU(2) \times U(1)$ gauge symmetry of the Standard Model.
\begin{table}
\begin{center}{\bf
Helicity Multiplets of $D=4$ Supergravities\\[5pt]

\begin{tabular}{|c|c|}
\hline
&\\[-3pt]
N & Helicity Content \\[6pt]
\hline
&\\[-3pt]
1& $\left[(2),\left(\frac{3}{2}\right)\right]$\\[6pt]
2& $\left[(2),2\left(\frac{3}{2}\right),(1)\right]$\\[6pt]
3& $\left[(2),3\left(\frac{3}{2}\right),3(1),\left(\frac{1}{2}\right)\right]$\\[6pt]
4& $\left[(2),4\left(\frac{3}{2}\right),6(1),4\left(\frac{1}{2}\right),2(0)\right]$\\[6pt]
5& $\left[(2),5\left(\frac{3}{2}\right),10(1),11\left(\frac{1}{2}\right),10(0)\right]$\\[6pt]
6& $\left[(2),6\left(\frac{3}{2}\right),16(1),26\left(\frac{1}{2}\right),30(0)\right]$\\[6pt]
8& $\left[(2),8\left(\frac{3}{2}\right),28(1),56\left(\frac{1}{2}\right),70(0)\right]$
\\[7pt]\hline
\end{tabular}}
\end{center}
\caption{The helicity content of supergravity multiplets in four dimensions. The factors indicate the multiplicities, and the overall multiplets for $N=7$ and $N=8$ coincide.} \label{tab:tab1}
\end{table}

\section{The Golden Age}
The following years, 1977 and 1978, were most performing, and a widespread enthusiasm drew into the field many new adepts. Important developments followed readily, including the discovery of ``minimal'' formulations where $N=1$ Supersymmetry is manifest \cite{auxiliary1,auxiliary2}. These involve extra (non-propagating) auxiliary degrees of freedom, which result into equal numbers of Bose and Fermi fields and particles. Finding these ``off-shell'' formulations has proved very hard beyond $N=1$, but in this case they have allowed systematic investigations of the spontaneous breaking of local Supersymmetry and precise characterizations of scalar geometries. This is fortunate, since only $N=1$ Supersymmetry is directly compatible with the chiral (parity-violating) interactions of the Standard Model. Auxiliary field formulations proved important also
to understand higher--derivative extensions of Supergravity \cite{higher_der_sugra}, including the supersymmetric extension \cite{susy+starobinsky} of Starobinsky's model of inflation \cite{starobinsky}. More recently, they were instrumental in supersymmetric localization techniques, where curved backgrounds providing infrared regulators are captured by diverse auxiliary--field configurations \cite{pestun}.

Complete matter couplings for $N=1,2$ in four dimensions were thus constructed by the early 1980’s \cite{complete1,complete2}, and some key features of the general $N=1$ case can be neatly summarized as in eq.~(\ref{eq3}) below:
\begin{eqnarray}
{\cal S} &=&  \int d^4 x \, e \,\biggl[  \frac{1}{2\, k^2} \ e^\mu_a\, e^\nu_b \, {R_{\mu\nu}}^{ab}(\omega) \ - \ \partial_i \, \partial_{\bar{j}}\,{\cal G} \, D_\mu z^i\, D_\nu {\bar{z}}^{\bar
{j}} \, g^{\mu\nu} \ - \  V(z,\bar {z})\ + \ \ldots \biggr] \ , \nonumber \\
&& V \ = \ e^{\cal G} \biggl[ {{\cal G}_{i}} \, {{\cal G}_{\bar j }} \ {\left({{\cal G}^{-1}}\right)^{i \bar j}} \ - \ 3 \biggr]\ , \qquad {\cal G}  \ = \ K \ + \ \log \left|W\right|^2 \ .\label{eq3}
\end{eqnarray}
The main ingredients of the construction are the Kahler potential $K(z^i,{\overline{z}}^{\overline i})$ and the superpotential $W(z^i)$, which depend on the scalar fields and enter the theory via the invariant combination ${\cal G}$.

A key step in the development of the theory had to do with the maximal model promptly identified by Gell-Mann. The maximal $N=8$ Supergravity was derived in 1978 by Eugene Cremmer and Bernard Julia~\cite{n8d4} from their previous, remarkable construction with Joel Scherk, of the unique Supergravity in eleven dimensions \cite{sugra11}. Its key features are summarized in
\begin{eqnarray}
{\cal S} &=& \frac{1}{2\, k^2}\
\int d^{11} x \, e\, \bigg[ e^\mu_a\, e^\nu_b \, {R_{\mu\nu}}^{ab}(\omega) \ - \ \overline{\psi}_\mu\, \gamma^{\mu\nu\rho}\, D_\nu\left(\frac{\omega+\hat{\omega}}{2}\right)\,\psi_\rho \nonumber \\
  &-& \frac{1}{24}\ F_{\mu\nu\rho\sigma}\,F^{\mu\nu\rho\sigma} \ - \ \frac{\sqrt{2}}{192} \ \left(\overline{\psi}_\mu \, \gamma^{\mu\nu\alpha\beta\gamma\delta} \, \psi_\nu \ + \ 12\, \overline{\psi}^\alpha \, \gamma^{\gamma\delta} \, \psi^\beta\right)\left(F_{\alpha\beta\gamma\delta} \ + \ {\widehat{F}}_{\alpha\beta\gamma\delta}\right) \nonumber \\
 &-& \frac{2\,\sqrt{2}}{(144)^2} \ \epsilon^{\alpha_1 \ldots \alpha_4 \beta_1 \ldots \beta_4 \mu\nu\rho}\ F^{\alpha_1\ldots \alpha_4}\ F^{\beta_1\ldots \beta_4}\, A_{\mu\nu\rho} \bigg] \ , \label{eq2} \\
&& {{\widehat{\omega}}_{\mu\,ab}} \ = \ {{\omega}_{\mu\,ab}} \ + \ \frac{1}{8} \ \overline{\psi}_\alpha\,{\gamma_{\mu a b}}^{\alpha\beta}\, \psi_\beta \ , \quad
 \widehat{F}_{\mu\nu\rho\sigma} \ = \
{F}_{\mu\nu\rho\sigma} \ + \ \frac{3\,\sqrt{2}}{2} \ \overline{\psi}_{[\mu}\, \gamma_{\nu\rho}\, \psi_{\sigma]} \ , \nonumber
\end{eqnarray}
where $\omega$ solves its field equation and ``hats'' denote supercovariant quantities.

Curiously, at most seven additional spatial dimensions are indeed allowed in Supergravity \cite{nahm}, in contrast with General Relativity.  In general, extra dimensions beyond our space time could exist, and yet be inaccessible to our senses, if they were curled into tiny internal spaces. This is the spirit of the Kaluza-Klein (KK) scenario that first linked, in the 1920’s, higher-dimensional GR and Electromagnetism \cite{kk}.

The first key ingredient of the $N=8$ construction was the geometrical nature of the scalar interactions, which result from an analogue of the pion model involving the 70 different fields present in the maximal theory, associated to the $E_{7(7)}/SU(8)$ coset. The second was a set of generalized electric-magnetic dualities, which extend the manifest symmetry of the vacuum Maxwell equations under the interchange of electric and magnetic fields and had already surfaced in simpler models \cite{dualities}. For instance, in the $N=8$ model, the $E_{7(7)}$ group acts on the 56 ``electric'' and ``magnetic'' field strengths as a generalized electric--magnetic duality~\cite{n8d4}. Hidden (infinite--dimensional) symmetries extending it have been widely explored in recent years, following  \cite{extended_exceptional}.

\section{Supergravity and Particle Physics}

 The MSSM \cite{mssm} and other supersymmetric extensions of the Standard Model were introduced and widely investigated by Pierre Fayet, Savas Dimopoulos, Howard Georgi and others \cite{mssm}, relying heavily on the soft-breaking terms proposed by Luciano Girardello and Marc Grisaru in \cite{girgri}. These low--dimensional couplings are not supersymmetric, and were introduced to overcome restrictions accompanying the spontaneous breaking of rigid Supersymmetry. A glimpse of the modifications induced by Supergravity is captured by the super--trace formula, here restricted for simplicity to chiral multiplets \cite{fgp,complete1},
 \begin{equation}
\left. \frac{1}{2} \ {\rm Str}\, M^2 \right|_{V_{z_i}=0,V=0} \ \equiv \ \frac{1}{2}\ \sum_J \ (-1)^{2\,J} \, (2\,J\,+\,1) \ m_J^2 \ = \ e^{\cal G} \biggl[ N\ -\ 1 \ + \ {\cal G}^i\, {{\cal G}^{\bar j}}\, {\cal R}_{i\bar j} \biggr] \ ,
\end{equation}
where $N$ is the number of scalar multiplets, which emerges in the $m_{3/2}/m_{Pl} \to 0$ limit. For the sake of comparison, in renormalizable models of rigid Supersymmetry the \emph{r.h.s.} would vanish, reflecting patterns where scalars pair, in mass, around fermions.

Supergravity provided a rationale for the emergence of soft--breaking terms as low--energy relics of the super--Higgs mechanism. This was shown by Riccardo Barbieri, Carlos Savoy and one of us (S.F.), and independently by Lawrence Hall, Joseph Lykken and Steven Weinberg, and by Richard Arnowitt, Ali Chamseddine and Pran Nath \cite{soft_susy}.

No-scale Supergravity, a theory with a naturally vanishing cosmological constant, was built in \cite{noscale}, and was readily applied to Physics beyond the Standard Model.

All these developments made it possible to derive parameter spaces of masses and couplings for supersymmetric extensions of the Standard Model, which can be explored in experimental searches for new physics, as was done at LEP, at Tevatron and elsewhere in the past and is currently done extensively at the LHC collider \cite{susy_developments}.

``Split Supersymmetry'' \cite{splitsusy} was advocated in recent years to obtain large mass separations among MSSM super partners while maintaining a number of attractive features of Supersymmetry, at the price of a fine--tuned cosmological constant \cite{weinberg_cosmo}, and was further developed by the same authors and many others.

\begin{table}
\begin{center}{\bf
Field content of $D=11$ and $D=10$ Supergravities and Super Yang--Mills\\[5pt]

\begin{tabular}{|c|c|}
\hline
&\\[-3pt]
Model & Field Content \\[6pt]
\hline
&\\[-3pt]
D=11 SGR& $\left(e_\mu^a,\psi_\mu,C_{\mu\nu\rho}\right)$\\[6pt]
\hline
&\\[-3pt]
D=10 IIA, or (1,1) SGR& $\left(e_\mu^a,\psi_{\mu\,L}, \psi_{\mu\,R},\lambda_{L}, \lambda_{R},C_{\mu\nu\rho},B_{\mu\nu},A_\mu,\phi\right)$\\[6pt]
D=10 IIB, or (2,0) SGR& $\left(e_\mu^a,\psi_{\mu\,L}^i, \lambda_{R}^i,{D^+}_{\mu\nu\rho\sigma},B_{\mu\nu}^i,\phi,\phi^\prime\right)\ \ \ \ (i=1,2)$\\[6pt]
D=10, (1,0) SGR& $\left(e_\mu^a,\psi_{\mu\,L},{\lambda}_{R},{B}_{\mu\nu},\phi\right)$\\[6pt]
\hline &\\[-3pt]
D=10, (1,0) SYM& $\left(A_\mu,\lambda_L\right)$
\\[6pt]\hline
\end{tabular}}
\end{center}
\caption{Standard field multiplets in $D=10,11$. The $IIA$ model is directly related via KK to eleven dimensions. The $IIB$ model contains doublets of gravitini, spinors and two-forms, and chiral four-form, $A^+$, with self--dual field strength, and two scalars parametrizing the $SL(2,R)/U(1)$ coset. The field content of the $I$ model is a truncation of the others. The $L,R$ suffixes indicate the chirality of Fermi fields.} \label{tab:tab2}
\end{table}

\section{Supergravity and String Theory}
The ultraviolet behavior of Supergravity theories was vigorously investigated soon after the original discovery. As in gravity \cite{thooft_veltman}, no divergences were found in the one-loop S-matrix of ``pure'' models involving solely the gravitational multiplet \cite{peter1loop}. Symmetry arguments soon pushed them at least to the three-loop order \cite{grisaru,deser}, and thus beyond GR, where divergences begin to show up at two loops \cite{goroff_as}. In subsequent years refined symmetry arguments and explicit computations relying on novel methods spurred by these studies have proceeded hand-in-hand, under the drive of Zvi Bern, Lance Dixon, David Kosower and others, revealing further, unexpected cancelations of divergences \cite{bern}.  The case of $N=8$ Supergravity remains unsettled, and some authors still envisage the possibility that this maximal theory be finite to all orders. In particular, the double--copy  structure relating $N=4$ Yang--Mills to $N=8$ Supergravity might provide a clue to its actual divergence structure \cite{bcj}, and recently played a role in the identification of an unexpected symmetry \cite{dual_conformal}, the dual superconformal symmetry, whose lessons could reverberate on Supergravity. Moreover, some subtleties that were well appreciated only recently might lead to deeper insights into the whole scenario \cite{bern_new}.

Supergravity allows in general continuous deformations that combine gauged internal symmetries and the emergence of scalar potentials. Explicit constructions soon clarified \cite{freedman_das,review_gs}, however, that these bring along negative vacuum energies, and thus maximally (super)symmetric anti-de Sitter, or AdS, vacua. This is the case, in particular, for the $N=8$ model of Bernard de Wit and Hermann Nicolai \cite{dewit_nicolai}, which realized the full $SO(8)$ gauge symmetry that we have already mentioned but does not admit a Minkowski vacuum. In a KK compactification from eleven dimensions, a gauged $SO(8)$ symmetry would be inherited from a seven-dimensional internal sphere \cite{seven_sphere}. A number of enticing variants were readily explored \cite{warner}.

The GSO projection opened the way to connecting String Theory to gravity along lines that were foreseen, in the mid 1970’s, by Scherk and Schwarz \cite{ss_74}, and independently by Tamiaki Yoneya \cite{yoneya}.  All ten-dimensional versions of Supergravity (and corresponding strings) were constructed by the 1980’s \cite{gs8084,gs80842}, but a widespread activity in this direction only started in 1984, when Michael Green and Schwarz discovered that gauge and gravitational \cite{agw} anomalies cancel, unexpectedly, in all versions of ten-dimensional supersymmetric String Theory (or Superstring Theory, for brief) \cite{gs84}. Anomalies are quantum violations of classical symmetries that are very troublesome when they affect gauge interactions. Their cancellation is a fundamental consistency condition, which is automatically granted in the Standard Model by its known particle content \cite{bim}.

The allowed gauge groups in ten dimensions are $SO(32)$ and $E_8 \times E_8$ in closed Heterotic superstrings \cite{ghmr}, whose discovery followed closely the anomaly cancellation, and $SO(32)$ in the Type-I theory involving both open and closed superstrings \cite{cp}, akin respectively to segments and circles. Here $E_8 \times E_8$ denotes two copies of the largest exceptional Lie algebra in the Cartan classification, and both $SO(32)$ and $E_8 \times E_8$ contain the $SU(3) \times SU(2) \times U(1)$ gauge symmetry of the Standard Model as a small subgroup. At low energies, all these theories reduce to minimal Supergravity coupled to a Yang-Mills multiplet in ten dimensions.

Unprecedented avenues thus opened up for linking ten-dimensional strings to the chiral interactions of Particle Physics, in enticing scenarios that are free, by construction, of the ultraviolet problems of gravity. One might well say that Supergravity started officially, with this 1984 ``first superstring revolution'', a second life as a low-energy manifestation of String Theory. It actually led to reconsider the very notion of anomaly, since in the Green-Schwarz mechanism some of its gauge potentials remove part of it. The original construction involved a single antisymmetric tensor potential $B_{\mu\nu}$ (to be contrasted with the electro-magnetic vector potential $A_\mu$), but more complicated cases that have emerged later rest on several fields of this and similar types.

The anomaly cancellation mechanism quickly resulted in definite KK scenarios granting String Theory a four-dimensional interpretation \cite{calabiyau}. It pointed to a specific class of six-dimensional internal manifolds leading to chiral spectra with $N=1$ Supersymmetry in four dimensions that were widely studied in Mathematics, Calabi-Yau spaces. These lead naturally, in this context, to a grand-unification gauge group $E_6$, which was known to connect to the Standard Model with right-handed neutrinos. All thus converged into an intriguing dictionary between the resulting low-energy dynamics and the topology of these complicated manifolds, or if you will between four-dimensional Supergravity and Algebraic Geometry. A remarkable property of these spaces, with deep physical implications, is the so-called ``mirror symmetry'' \cite{yau}, which exchanges manifolds while swapping their Hodge numbers. This symmetry allows one to compute non--perturbative couplings in terms of classical ones, and finds a rationale in the $C$-map of the four--dimensional low--energy Supergravity \cite{cmap}, where $N=2$ vector multiplets and hypermultiplets are interchanged. The structure of these multiplets rests on ``special geometry'', whose mathematical significance was first stressed in \cite{strominger}. No-scale Supergravity \cite{noscale} was linked to Calabi-Yau compactifications in \cite{witten85,gkp}.
Actually, strings behave properly also on singular (orbifold) limits that compare to smooth Calabi-Yau spaces like tetrahedra to spheres \cite{orbifolds}. These limits afford more complete descriptions, and have led to deep insights into the structure of String Theory, opening the way to early direct constructions of four--dimensional heterotic string spectra \cite{closed_4d}

Other (type II) theories with maximal Supersymmetry exist in ten dimensions, are also free of anomalies and allow similar compactifications, but for a while they seemed totally unrelated to Particle Physics.

\section{Branes and M--Theory}
The early 1990’s were marked by many detailed studies of classical solutions of Supergravity \cite{classical_solutions}.  These generalize BH's, and form fields are their key new ingredient. In four dimensions, electric charges source a vector potential $A_\mu$, whose Lorentz index reflects somehow their ``world lines''.  In a similar fashion, strings source antisymmetric potentials $B_{\mu\nu}$, while other potentials present in ten-dimensional Supergravity are related to extended objects of higher dimensionality, generically dubbed ``$p$-branes''. Here $p$ is the number of spatial dimensions of these extended objects ($p=0$ for particles, $p=1$ for strings, $p=2$ for membranes, and so on). Supergravity cannot forego the presence of a wide class of extended objects, with which String Theory would appear at odds, insofar as it focuses solely on strings. This was strongly advocated, over the years, by a number of scientists, and most notably by Michael Duff \cite{mjd} and Paul Townsend \cite{pkt}.
\begin{figure}[ht]
\centering
\includegraphics[width=8cm]{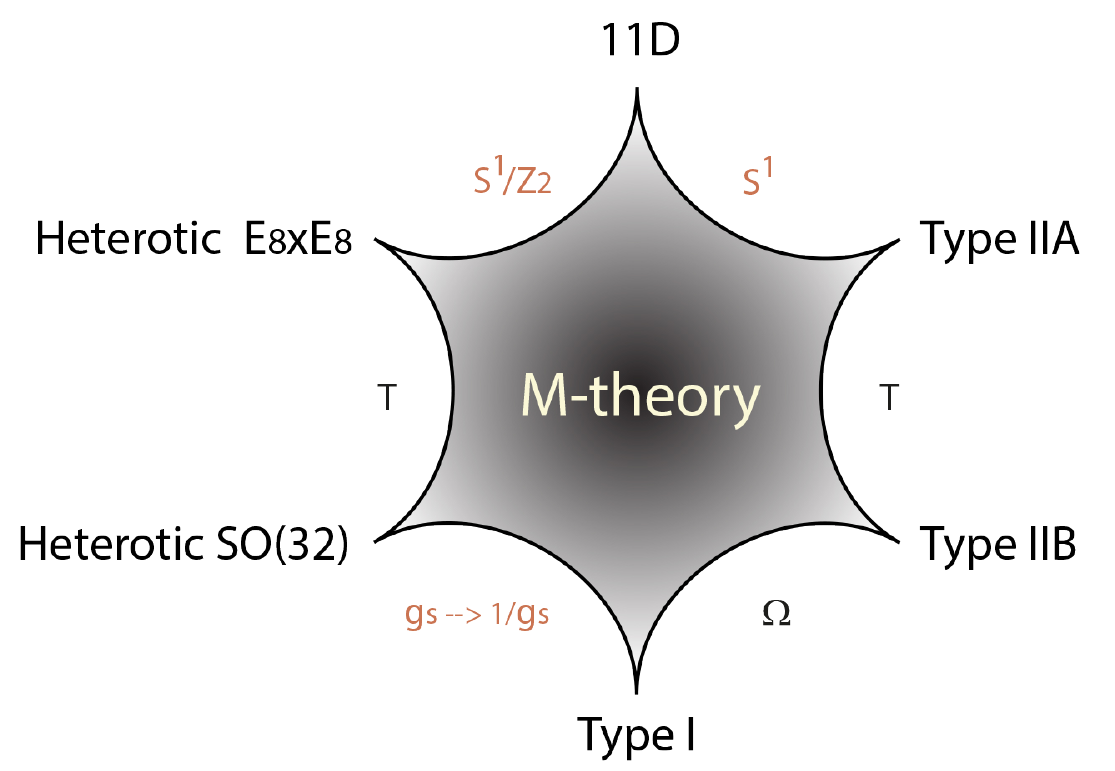}
\caption{Duality links among the five ten--dimensional supersymmetric versions of String Theory and eleven--dimensional Supergravity.}
\label{fig:hexagon}
\end{figure}

The ``second superstring revolution'' is associated to the mid-1990’s, and to the resolution of this apparent dichotomy when it became widely recognized that String Theory, after all, cannot be merely a theory of strings. Behind it, as we now concede, lies indeed a far more complicated soup of strings and more general $p$-branes. The novel ingredient of this wider picture was a class of $p$-branes that show up in string perturbation theory in the presence of boundaries, the D-branes whose role was clarified by Joseph Polchinski \cite{jp}, but the (electric-magnetic) dualities of the low-energy Supergravity were again a main tool. Strong--weak coupling dualities in String Theory were actually first advocated by Anna-Maria Font, Dieter Lust, Luis Ibanez and Fernando Quevedo in \cite{fliq}, and were inspired by continuous symmetries of the low--energy Supergravity. The end result is an awesome unified picture, which was largely due to Edward Witten \cite{witten95} and is usually referred to as M-theory.

When strings combine with other extended objects, even the very notion of space time is blurred, to the extent that ten and eleven dimensions emerge from singular corners of this correspondence. One of these is the eleven-dimensional Supergravity. These results are usually summarized via the hexagon--like diagram of Fig.~\ref{fig:hexagon}, whose sides reflect different duality links, three of which were deeply inspired by Supergravity \cite{ht,witten95,horwit}. The others had already surfaced in the late Eighties: they are beyond its reach but find their rationale in T-duality, a peculiar string correspondence between large and small KK radii \cite{Tduality}, and in the $\Omega$ orientifold link between type-IIB and type-I strings proposed by one of us (A.~S.) \cite{orientifolds}. Supergravity thus started officially a third, parallel life, as a probe into the elusive inner workings of String Theory.~\footnote{Recently, double field theory \cite{double_ft} has emerged as an enticing intermediate framework to accommodate T-duality within Field Theory.}

\section{Supergravity and the AdS/CFT Correspondence}
The late 1990’s witnessed the emergence of a peculiar duality that has had a huge impact on the literature. This is the AdS/CFT correspondence \cite{adscft}, which links ``holographically'' gravity and gauge theories~\footnote{There were a number of notable anticipations of aspects of the correspondence, including some contributions in \cite{gs80842,anticipations}.}. It was originally conjectured by Juan Maldacena, and its most explored form concerns String Theory compactified on a special KK manifold, $AdS_5 \times S_5$, the direct product of a five-dimensional anti-de Sitter space and a five-dimensional internal sphere. $AdS_5$ has a boundary at spatial infinity, which is identified with a four-dimensional Minkowski space, and the correspondence posits the equivalence between a weakly coupled String Theory in the bulk and a strongly coupled field theory, the $N=4$ supersymmetric Yang-Mills theory that we have already met in a ten--dimensional guise, on this boundary. This surprising correspondence, which brought AdS Supergravity to the forefront, also vindicates some intriguing ideas that had been around since Jacob Bekenstein connected the black-hole entropy to its area \cite{holography}. Supergravity thus started officially, with this ``third superstring revolution'', one more parallel life as an unprecedented tool for exploring non-perturbative features of gauge theories.

In $N=2$ Supergravity supersymmetric black holes give rise to universal (duality--invariant) area formulas for their Bekenstein-Hawking Entropy, which generalize the Reissner--Nordstrom entropy formula for charged dyons and rest on the (square modulus) of the central charge computed at its extremum in moduli space \cite{attractors}. The name ``attractors'' was coined since, independently of the initial conditions at large distances, the scalar trajectories in moduli space approach the same extremal points, which coincide with the values at the BH Horizon. This explains, in particular, why the ADM mass is a continuous moduli dependent function while the entropy is quantized in terms of BH charges. In the large--charge limit, Andrew Strominger and Cumrun Vafa associated to these types of results for extremal BH's a microscopic String Theory counting \cite{strom_vafa}. The actual nature of the microstates, however, is still a debated issue \cite{microstates}.

\section{Conclusions and Perspectives}

The last two decades have witnessed a multitude of applications of AdS/CFT outside its original realm. These have touched upon fluid dynamics, the quark-gluon plasma, and more recently Condensed Matter Physics, providing a number of useful insights on strongly coupled matter systems. Perhaps more unexpectedly, AdS/CFT duality has stimulated work related to scattering amplitudes, which may also shed light on the old issue of the ultraviolet behavior of supergravity, but the reverse program of gaining information about gravity from gauge dynamics has proved harder. Above all, however, there is a pressing need to shed light on the geometrical principles and the deep symmetries underlying String Theory, which have proved elusive over the years.

The interplay between Particle Physics and Cosmology is a natural arena to explore consequences of Supergravity. Recent experiments probing the Cosmic Microwave Background, and in particular the results of the Planck mission \cite{planck}, have lent some definite support to inflationary models of the Early Universe. An elusive particle, the inflaton, should have driven this primordial acceleration, and while our current grasp of String Theory does not allow a detailed analysis of the problem, Supergravity can provide fundamental clues on this and the subsequent particle physics epochs.

Applications of Supergravity to Cosmology have attracted an increasing interest in recent years \cite{appl_cosmology,susy+starobinsky}. Supersymmetry is broken in a de Sitter like inflationary phase, where one typically encounters more fields than would be needed for the early Universe, and moreover some familiar scenarios tend to be plagued by instabilities. The novel ingredient that appears to get around these problems is non-linear supersymmetry \cite{cosmo_nl}, whose foundations lie in the prescient 1973 work of Volkov and Akulov \cite{va} and on the technique of constrained superfields \cite{constrained_superfields}. Non-linear supersymmetry arises when some super-partners are exceedingly massive, seems to play an intriguing role in String Theory \cite{bsb} and connects naturally to the KKLT scenario \cite{KKLT}. The current lack of signals for supersymmetry at the LHC makes one wonder whether it might also hold a prominent place in an eventual picture of Particle Physics. This resonates with the idea of ``split supersymmetry'' \cite{splitsusy}, which allows for large mass separations among superpartners and can be accommodated in Supergravity at the price of reconsidering hierarchy issues.

Crossing the current frontiers appears to require a deeper understanding of broken Supersymmetry in Supergravity and in String Theory. Broken Supersymmetry made an early entry in String Theory via orbifold realizations of the Scherk-Schwarz mechanism \cite{ss} in models of oriented closed strings \cite{ss_closed}. The resulting spectra are closely connected to special versions of gauged Supergravity, whose AdS versions play also a central role in the AdS/CFT correspondence, and afford interesting generalizations in String Theory in the presence of branes \cite{ss_branes}, even in the presence of internal magnetic fields \cite{magnetic}. These constructions involve the Born--Infeld action \cite{borninfeld}, which was originally recovered from open strings in \cite{fradkin_tseytlin}. Branes were actually instrumental in Supersymmetry in another context, since they provided early clues on mechanisms for its partial breaking \cite{hp}, and this phenomenon is connected again, in the non--linear limit, to extensions of the Born--Infeld theory \cite{bi_nlin}. Finally, in recent years, gauged Supergravity found a geometrical framework within generalized geometry \cite{gen_geom}, a setting that is related to the flux compactifications of \cite{gkp}, while the embedding tensor of \cite{embedding,review_gs} captures their algebraic foundations.

In breaking Supersymmetry, one is confronted with important conceptual challenges: the resulting vacua are deeply affected, in general, by quantum fluctuations, and this reverberates on old conundrums related to dark energy and the cosmological constant. There are even signs that this type of investigations could shed light on the backbone of String Theory, and we are confident that Supergravity will lead us farther once more. Finally, Supergravity may have something to say about the dark matter in the Universe, since gravitini or other light superpartners dubbed neutralinos might perhaps account for it.

\vskip 24pt

\subsection*{Acknowledgements}

This overview was based in part on the talk delivered by S.~F. at the ``Infeld Colloquium and Discrete'', in Warsaw, on December 1 2016, and on a joint CERN Courier article \cite{cerncourier}. It is focussed on topics that are closer to the research interests of the authors, who apologize in advance if they have missed some important references. SF was supported in part by the CERN TH-Department and by INFN (IS CSN4-GSS-PI). AS was supported in part by Scuola Normale, by INFN (IS CSN4-GSS-PI) and by APC-U.~Paris VII, and would like to thank APC-U.~Paris VII for the kind hospitality. We are grateful to Carlo Angelantonj, Bianca Letizia Cerchiai, Emilian Dudas and Marine Samsonyan, who read the manuscript and offered constructive suggestions.


\newpage
\begin{thebibliography}{999}
\bibitem{sugra} For a recent review see: D.~Z.~Freedman and A.~Van Proeyen,
  ``Supergravity,''
  Cambridge, UK: Cambridge Univ. Pr. (2012). For an early review see:
  P.~Van Nieuwenhuizen,
  Phys.\ Rept.\  {\bf 68} (1981) 189.

\bibitem{gr} For reviews see: C.~W.~Misner, K.~S.~Thorne and J.~A.~Wheeler,  ``Gravitation,''  (Freeman, 1970);S.~Weinberg,
  ``Gravitation and Cosmology : Principles and Applications of the General Theory of Relativity,'' (Wiley, 1972)
  S.~W.~Hawking and G.~F.~R.~Ellis,
  ``The Large Scale Structure of Space-Time,'' (Cambridge Univ. Press, 1973);
  R.~M.~Wald,
  ``General Relativity,''
  (Chicago Univ. Pr., 1984)

\bibitem{susy}  J.~Wess and B.~Zumino,
  Phys.\ Lett.\  {\bf 49B} (1974) 52,
  Nucl.\ Phys.\ B {\bf 70} (1974) 39;
A.~Salam and J.~A.~Strathdee,
  Nucl.\ Phys.\ B {\bf 76} (1974) 477;
  S.~Ferrara and B.~Zumino,
  Nucl.\ Phys.\ B {\bf 79} (1974) 413;
  A.~Salam and J.~A.~Strathdee,
  Phys.\ Lett.\  {\bf 51B} (1974) 353;
  P.~Fayet and J.~Iliopoulos,
  Phys.\ Lett.\ B {\bf 51} (1974) 461.

\bibitem{susy_developments}
 For a review of early results see:
  P.~Fayet and S.~Ferrara,
  Phys.\ Rept.\  {\bf 32} (1977) 249.
  For a collection of early papers see:
  S.~Ferrara,
  ``Supersymmetry,''
  Amsterdam, Netherlands: North-Holland (1987) 1343 p., Singapore, Singapore: World Scientific (1987) 1343p.
  More recent reviews are:
  P.~C.~West,
  ``Introduction to supersymmetry and supergravity,''
  Singapore, Singapore: World Scientific (1990) 425 p
J.~Wess and J.~Bagger,
  ``Supersymmetry and supergravity,''
  Princeton, USA: Univ. Pr. (1992) 259 p.;
 S.~Weinberg,
  ``The quantum theory of fields. Vol. 3: Supersymmetry,''
  ISBN-9780521670555.


\bibitem{cm} S.~R.~Coleman and J.~Mandula,
  Phys.\ Rev.\  {\bf 159} (1967) 1251;
  R.~Haag, J.~T.~Lopuszanski and M.~Sohnius,
  Nucl.\ Phys.\ B {\bf 88} (1975) 257.

\bibitem{ven} G.~Veneziano,
  Nuovo Cim.\ A {\bf 57} (1968) 190.

\bibitem{nsr}A.~Neveu and J.~H.~Schwarz,
  Nucl.\ Phys.\ B {\bf 31} (1971) 86;
  P.~Ramond,
  Phys.\ Rev.\ D {\bf 3} (1971) 2415.

\bibitem{gs} J.~L.~Gervais and B.~Sakita,
  Nucl.\ Phys.\ B {\bf 34} (1971) 632.

\bibitem{gl} Y.~A.~Golfand and E.~P.~Likhtman,
  JETP Lett.\  {\bf 13} (1971) 323
   [Pisma Zh.\ Eksp.\ Teor.\ Fiz.\  {\bf 13} (1971) 452].

\bibitem{va} D.~V.~Volkov and V.~P.~Akulov,
  Phys.\ Lett.\  {\bf 46B} (1973) 109.

\bibitem{beh} F.~Englert and R.~Brout,
  Phys.\ Rev.\ Lett.\  {\bf 13} (1964) 321;
  P.~W.~Higgs,
  Phys.\ Lett.\  {\bf 12} (1964) 132,
  Phys.\ Rev.\ Lett.\  {\bf 13} (1964) 508,
  Phys.\ Rev.\  {\bf 145} (1966) 1156.

\bibitem{lhc} G.~Aad {\it et al.} [ATLAS Collaboration],
  Phys.\ Lett.\ B {\bf 716} (2012) 1
  [arXiv:1207.7214 [hep-ex]];
  S.~Chatrchyan {\it et al.} [CMS Collaboration],
  Phys.\ Lett.\ B {\bf 716} (2012) 30
  [arXiv:1207.7235 [hep-ex]].

\bibitem{mssm} P.~Fayet,
  Phys.\ Lett.\  {\bf 69B} (1977) 489;
S.~Dimopoulos and H.~Georgi,
  Nucl.\ Phys.\ B {\bf 193} (1981) 150,
and references therein.
  For a review see:
  H.~P.~Nilles,
  Phys.\ Rept.\  {\bf 110} (1984) 1.


\bibitem{cosmology} For reviews see: V.~Mukhanov,
  ``Physical foundations of cosmology,''
  Cambridge, UK: Univ. Pr. (2005);
S.~Weinberg, ``Cosmology,''
 Oxford, UK: Oxford Univ. Pr. (2008);
D.~H.~Lyth and A.~R.~Liddle,
 ``The primordial density perturbation: Cosmology, inflation and the origin of structure,''
  Cambridge, UK: Cambridge Univ. Pr. (2009);
  D. S. Gorbunov and V. A. Rubakov, ``Introduction
to the theory of the early Universe: Cosmological perturbations and inflationary
theory,'' Hackensack, USA: World Scientific (2011).

\bibitem{inflation}
 A.~A.~Starobinsky,
  Phys.\ Lett.\ {\bf B 91} (1980) 99;
  D.~Kazanas,
  Astrophys.\ J.\  {\bf 241} (1980) L59;
  K.~Sato,
  Phys.\ Lett.\ B {\bf 99} (1981) 66;
 A.~H.~Guth,
 Phys.\ Rev.\ {\bf D 23} (1981) 347;
   V.~F.~Mukhanov and G.~V.~Chibisov,
  JETP Lett.\  {\bf 33} (1981) 532
   [Pisma Zh.\ Eksp.\ Teor.\ Fiz.\  {\bf 33} (1981) 549];
 A.~D.~Linde,
  Phys.\ Lett.\ {\bf B 108} (1982) 389;
   A.~Albrecht and P.~J.~Steinhardt,
  Phys.\ Rev.\ Lett.\  {\bf 48} (1982) 1220;
   A.~D.~Linde,
  Phys.\ Lett.\ {\bf B129} (1983) 177.
 For reviews see:
 N.~Bartolo, E.~Komatsu, S.~Matarrese and A.~Riotto,
  Phys.\ Rept.\  {\bf 402} (2004) 103
  [astro-ph/0406398];
   J.~Martin, C.~Ringeval and V.~Vennin,
  Phys.\ Dark Univ.\  {\bf 5-6} (2014) 75
  [arXiv:1303.3787 [astro-ph.CO]].

\bibitem{gravitywaves} B.~P.~Abbott {\it et al.} [LIGO Scientific and Virgo Collaborations],
  Phys.\ Rev.\ Lett.\  {\bf 116} (2016) no.6,  061102
  [arXiv:1602.03837 [gr-qc]].

\bibitem{ffvn} D.~Z.~Freedman, P.~van Nieuwenhuizen and S.~Ferrara,
  Phys.\ Rev.\ D {\bf 13} (1976) 3214.

\bibitem{dz} S.~Deser and B.~Zumino,
  Phys.\ Lett.\  {\bf 62B} (1976) 335.

\bibitem{vz}
G.~Velo and D.~Zwanziger,
  Phys.\ Rev.\  {\bf 188} (1969) 2218.

\bibitem{higher_spins}
A.~Sagnotti,
  J.\ Phys.\ A {\bf 46} (2013) 214006
  [arXiv:1112.4285 [hep-th]].
 and the other papers in the same special issue of J. Phys. A devoted to higher spins.

\bibitem{mmans}
S.~W.~MacDowell and F.~Mansouri,
  Phys.\ Rev.\ Lett.\  {\bf 38} (1977) 739
   Erratum: [Phys.\ Rev.\ Lett.\  {\bf 38} (1977) 1376].

\bibitem{torino}
Y.~Ne'eman and T.~Regge,
  Phys.\ Lett.\  {\bf 74B} (1978) 54.
  For a review see:
  L.~Castellani, R.~D'Auria and P.~Fr\'e,
  ``Supergravity and superstrings'' , 3 vols
  Singapore, Singapore: World Scientific (1991).

\bibitem{broken_superconformal}
M.~Kaku, P.~K.~Townsend and P.~van Nieuwenhuizen,
  Phys.\ Lett.\  {\bf 69B} (1977) 304.

\bibitem{firstmattersugra} S.~Ferrara, D.~Z.~Freedman, P.~van Nieuwenhuizen, P.~Breitenlohner, F.~Gliozzi and J.~Scherk,
  Phys.\ Rev.\ D {\bf 15} (1977) 1013;
  S.~Ferrara, F.~Gliozzi, J.~Scherk and P.~Van Nieuwenhuizen,
  Nucl.\ Phys.\ B {\bf 117} (1976) 333.

\bibitem{strings} For reviews see: M.~B.~Green, J.~H.~Schwarz and E.~Witten, ``Superstring Theory'', 2 vols., Cambridge, UK: Cambridge Univ. Press (1987); J.~Polchinski, ``String theory'', 2 vols. Cambridge, UK: Cambridge Univ. Press (1998);  C.~V.~Johnson, ``D-branes,'' USA: Cambridge Univ. Press (2003) 548 p; B.~Zwiebach, ``A first course in string theory''
Cambridge, UK: Cambridge Univ. Press (2004); K.~Becker, M.~Becker and J.~H.~Schwarz,
``String theory and M-theory: A modern introduction'' Cambridge, UK: Cambridge Univ.
Press (2007); E.~Kiritsis, ``String theory in a nutshell'', Princeton, NJ: Princeton Univ. Press (2007);
P.~West, ``Introduction to strings and branes,'' Cambridge: Cambridge Univ. Press (2012).

\bibitem{gso} F.~Gliozzi, J.~Scherk and D.~I.~Olive,
  Nucl.\ Phys.\ B {\bf 122} (1977) 253.

\bibitem{n2sugra} S.~Ferrara and P.~van Nieuwenhuizen,
  Phys.\ Rev.\ Lett.\  {\bf 37} (1976) 1669.

\bibitem{su4_sugra}
E.~Cremmer, J.~Scherk and S.~Ferrara,
  Phys.\ Lett.\  {\bf 74B} (1978) 61.

\bibitem{n4ym} L.~Brink, J.~H.~Schwarz and J.~Scherk,
  Nucl.\ Phys.\ B {\bf 121} (1977) 77.

\bibitem{grs} M.~Gell-Mann, P.~Ramond and R.~Slansky,
  Rev.\ Mod.\ Phys.\  {\bf 50} (1978) 721.

\bibitem{auxiliary1} S.~Ferrara and P.~van Nieuwenhuizen,
  Phys.\ Lett.\  {\bf 74B} (1978) 333;
  K.~S.~Stelle and P.~C.~West,
  Phys.\ Lett.\  {\bf 74B} (1978) 330.

\bibitem{auxiliary2}  M.~F.~Sohnius and P.~C.~West,
  Phys.\ Lett.\  {\bf 105B} (1981) 353,
  Nucl.\ Phys.\ B {\bf 198} (1982) 493;
  S.~Ferrara and S.~Sabharwal,
  Annals Phys.\  {\bf 189} (1989) 318.

\bibitem{higher_der_sugra}
S.~Ferrara, M.~T.~Grisaru and P.~van Nieuwenhuizen,
  Nucl.\ Phys.\ B {\bf 138} (1978) 430;
  S.~Cecotti,
  Phys.\ Lett.\ B {\bf 190} (1987) 86.

\bibitem{susy+starobinsky}
F.~Farakos, A.~Kehagias and A.~Riotto,
  Nucl.\ Phys.\ B {\bf 876} (2013) 187
  [arXiv:1307.1137 [hep-th]];
S.~Ferrara, R.~Kallosh, A.~Linde and M.~Porrati,
  Phys.\ Rev.\ D {\bf 88} (2013) no.8,  085038
  [arXiv:1307.7696 [hep-th]].

\bibitem{starobinsky}
A.~A.~Starobinsky,
  Phys.\ Lett.\  {\bf 91B} (1980) 99.

\bibitem{pestun}
V.~Pestun,
  Commun.\ Math.\ Phys.\  {\bf 313} (2012) 71
  [arXiv:0712.2824 [hep-th]];
  G.~Festuccia and N.~Seiberg,
  JHEP {\bf 1106} (2011) 114
  [arXiv:1105.0689 [hep-th]];
    F.~Benini and A.~Zaffaroni,
  arXiv:1605.06120 [hep-th].

\bibitem{complete1} E.~Cremmer, B.~Julia, J.~Scherk, S.~Ferrara, L.~Girardello and P.~van Nieuwenhuizen,
  Nucl.\ Phys.\ B {\bf 147} (1979) 105;
  J.~Bagger and E.~Witten,
  Phys.\ Lett.\  {\bf 118B} (1982) 103;
  E.~Cremmer, S.~Ferrara, L.~Girardello and A.~Van Proeyen,
  Nucl.\ Phys.\ B {\bf 212} (1983) 413.

\bibitem{complete2} M.~Gunaydin, G.~Sierra and P.~K.~Townsend,
  Nucl.\ Phys.\ B {\bf 242} (1984) 244;
  J.~Bagger and E.~Witten,
  Nucl.\ Phys.\ B {\bf 222} (1983) 1;
  B.~de Wit, P.~G.~Lauwers and A.~Van Proeyen,
  Nucl.\ Phys.\ B {\bf 255} (1985) 569;
    E.~Cremmer, C.~Kounnas, A.~Van Proeyen, J.~P.~Derendinger, S.~Ferrara, B.~de Wit and L.~Girardello,
  Nucl.\ Phys.\ B {\bf 250} (1985) 385;
  L.~Andrianopoli, M.~Bertolini, A.~Ceresole, R.~D'Auria, S.~Ferrara, P.~Fr\'e and T.~Magri,
  J.\ Geom.\ Phys.\  {\bf 23} (1997) 111
  [hep-th/9605032].
  For a review see:
  H.~Samtleben,
  Class.\ Quant.\ Grav.\  {\bf 25} (2008) 214002
  [arXiv:0808.4076 [hep-th]].
For a review of harmonic Supersace see:
A.~S.~Galperin, E.~A.~Ivanov, V.~I.~Ogievetsky and E.~S.~Sokatchev,
  ``Harmonic Superspace,'' Cambridge, UK: Cambridge Univ. Pr. (2001).

\bibitem{n8d4} E.~Cremmer and B.~Julia,
  Nucl.\ Phys.\ B {\bf 159} (1979) 141.

\bibitem{sugra11} E.~Cremmer, B.~Julia and J.~Scherk,
  Phys.\ Lett.\  {\bf 76B} (1978) 409.

\bibitem{nahm} W.~Nahm,
  Nucl.\ Phys.\ B {\bf 135} (1978) 149.

\bibitem{kk} T.~Kaluza,
  Sitzungsber.\ Preuss.\ Akad.\ Wiss.\ Berlin (Math.\ Phys.) {\bf 1921} (1921) 966;
  O.~Klein,
  Z.\ Phys.\  {\bf 37} (1926) 895
   [Surveys High Energ.\ Phys.\  {\bf 5} (1986) 241].

\bibitem{dualities} S.~Ferrara, J.~Scherk and B.~Zumino,
  Nucl.\ Phys.\ B {\bf 121} (1977) 393;
  M.~K.~Gaillard and B.~Zumino,
  Nucl.\ Phys.\ B {\bf 193} (1981) 221.
  For a review see:
  P.~Aschieri, S.~Ferrara and B.~Zumino,
  Riv.\ Nuovo Cim.\  {\bf 31} (2008) 625
  [arXiv:0807.4039 [hep-th]].

\bibitem{extended_exceptional}
P.~C.~West,
  Class.\ Quant.\ Grav.\  {\bf 18} (2001) 4443
  [hep-th/0104081];
T.~Damour, M.~Henneaux and H.~Nicolai,
  Phys.\ Rev.\ Lett.\  {\bf 89} (2002) 221601
  [hep-th/0207267].

\bibitem{girgri}
L.~Girardello and M.~T.~Grisaru,
  Nucl.\ Phys.\ B {\bf 194} (1982) 65.

\bibitem{fgp}
S.~Ferrara, L.~Girardello and F.~Palumbo,
  Phys.\ Rev.\ D {\bf 20} (1979) 403;
  M.~T.~Grisaru, M.~Rocek and A.~Karlhede,
  Phys.\ Lett.\  {\bf 120B} (1983) 110.

\bibitem{soft_susy}
R.~Barbieri, S.~Ferrara and C.~A.~Savoy,
  Phys.\ Lett.\  {\bf 119B} (1982) 343;
A.~H.~Chamseddine, R.~L.~Arnowitt and P.~Nath,
  Phys.\ Rev.\ Lett.\  {\bf 49} (1982) 970;
L.~J.~Hall, J.~D.~Lykken and S.~Weinberg,
  Phys.\ Rev.\ D {\bf 27} (1983) 2359;
  G.~F.~Giudice and A.~Masiero,
  Phys.\ Lett.\ B {\bf 206} (1988) 480.

\bibitem{noscale}
E.~Cremmer, S.~Ferrara, C.~Kounnas and D.~V.~Nanopoulos,
  Phys.\ Lett.\  {\bf 133B} (1983) 61;
J.~R.~Ellis, A.~B.~Lahanas, D.~V.~Nanopoulos and K.~Tamvakis,
  Phys.\ Lett.\  {\bf 134B} (1984) 429.
  For a review see:
  A.~B.~Lahanas and D.~V.~Nanopoulos,
  Phys.\ Rept.\  {\bf 145} (1987) 1.

\bibitem{splitsusy}
 J.~D.~Wells,
  hep-ph/0306127;
  N.~Arkani-Hamed and S.~Dimopoulos,
  JHEP {\bf 0506} (2005) 073
  [hep-th/0405159];
  G.~F.~Giudice and A.~Romanino,
  Nucl.\ Phys.\ B {\bf 699} (2004) 65
   Erratum: [Nucl.\ Phys.\ B {\bf 706} (2005) 487]
  [hep-ph/0406088];
  N.~Arkani-Hamed, S.~Dimopoulos, G.~F.~Giudice and A.~Romanino,
  Nucl.\ Phys.\ B {\bf 709} (2005) 3
  [hep-ph/0409232].

\bibitem{weinberg_cosmo}
For a classic review, see:
S.~Weinberg,
  Rev.\ Mod.\ Phys.\  {\bf 61} (1989) 1.

\bibitem{thooft_veltman} G.~'t Hooft and M.~J.~G.~Veltman,
  Ann.\ Inst.\ H.\ Poincare Phys.\ Theor.\ A {\bf 20} (1974) 69.

\bibitem{peter1loop} M.~T.~Grisaru, P.~van Nieuwenhuizen and J.~A.~M.~Vermaseren,
  Phys.\ Rev.\ Lett.\  {\bf 37} (1976) 1662.

\bibitem{grisaru} M.~T.~Grisaru,
  Phys.\ Lett.\  {\bf 66B} (1977) 75;
  E.~Tomboulis,
  Phys.\ Lett.\  {\bf 67B} (1977) 417.

\bibitem{deser}
  S.~Deser, J.~H.~Kay and K.~S.~Stelle,
  Phys.\ Rev.\ Lett.\  {\bf 38} (1977) 527
  [arXiv:1506.03757 [hep-th]];
  R.~E.~Kallosh,
  Phys.\ Lett.\  {\bf 99B} (1981) 122,
  M.~T.~Grisaru and W.~Siegel,
  Nucl.\ Phys.\ B {\bf 187} (1981) 149;
  Nucl.\ Phys.\ B {\bf 201} (1982) 292
   Erratum: [Nucl.\ Phys.\ B {\bf 206} (1982) 496].

\bibitem{goroff_as} M.~H.~Goroff and A.~Sagnotti,
  Phys.\ Lett.\  {\bf 160B} (1985) 81,
  Nucl.\ Phys.\ B {\bf 266} (1986) 709;
  A.~E.~M.~van de Ven,
  Nucl.\ Phys.\ B {\bf 378} (1992) 309.

\bibitem{bern} For recent reviews see:
K.~S.~Stelle,
  Int.\ J.\ Geom.\ Meth.\ Mod.\ Phys.\  {\bf 09} (2012) 1261013;
Z.~Bern,
  Mod.\ Phys.\ Lett.\ A {\bf 29} (2014) no.32,  1430036.
  For a comprehensive discussion of superfield methods see:
 S.~J.~Gates, M.~T.~Grisaru, M.~Rocek and W.~Siegel,
  ``Superspace Or One Thousand and One Lessons in Supersymmetry,''
  Front.\ Phys.\  {\bf 58} (1983) 1
  [hep-th/0108200].
\bibitem{bcj}
Z.~Bern, J.~J.~M.~Carrasco and H.~Johansson,
  Phys.\ Rev.\ Lett.\  {\bf 105} (2010) 061602
  [arXiv:1004.0476 [hep-th]].

\bibitem{dual_conformal}
J.~M.~Drummond, J.~Henn, G.~P.~Korchemsky and E.~Sokatchev,
  Nucl.\ Phys.\ B {\bf 828} (2010) 317
  [arXiv:0807.1095 [hep-th]].

\bibitem{bern_new}
Z.~Bern, C.~Cheung, H.~H.~Chi, S.~Davies, L.~Dixon and J.~Nohle,
  Phys.\ Rev.\ Lett.\  {\bf 115} (2015) no.21,  211301
  [arXiv:1507.06118 [hep-th]];
Z.~Bern, H.~H.~Chi, L.~Dixon and A.~Edison,
  arXiv:1701.02422 [hep-th].

\bibitem{freedman_das}D.~Z.~Freedman and A.~K.~Das,
  Nucl.\ Phys.\ B {\bf 120} (1977) 221.

\bibitem{review_gs}
For a recent review, see:
M.~Trigiante,
  arXiv:1609.09745 [hep-th].

\bibitem{dewit_nicolai} B.~de Wit and H.~Nicolai,
  Nucl.\ Phys.\ B {\bf 208} (1982) 323.

\bibitem{seven_sphere}  B.~de Wit and H.~Nicolai,
  Nucl.\ Phys.\ B {\bf 281} (1987) 211.

\bibitem{warner}
C.~M.~Hull,
  Phys.\ Lett.\  {\bf 142B} (1984) 39;
M.~Gunaydin, L.~J.~Romans and N.~P.~Warner,
  Phys.\ Lett.\  {\bf 154B} (1985) 268,
  Nucl.\ Phys.\ B {\bf 272} (1986) 598.

\bibitem{ss_74}  J.~Scherk and J.~H.~Schwarz,
  Nucl.\ Phys.\ B {\bf 81} (1974) 118.

\bibitem{yoneya} T.~Yoneya,
  Prog.\ Theor.\ Phys.\  {\bf 51} (1974) 1907.

\bibitem{gs8084} M.~B.~Green and J.~H.~Schwarz,
  Phys.\ Lett.\  {\bf 109B} (1982) 444;
  M.~B.~Green, J.~H.~Schwarz and L.~Brink,
  Nucl.\ Phys.\ B {\bf 198} (1982) 474;
  J.~H.~Schwarz and P.~C.~West,
  Phys.\ Lett.\  {\bf 126B} (1983) 301;
  P.~S.~Howe and P.~C.~West,
  Nucl.\ Phys.\ B {\bf 238} (1984) 181;
  E.~Bergshoeff, M.~de Roo, B.~de Wit and P.~van Nieuwenhuizen,
  Nucl.\ Phys.\ B {\bf 195} (1982) 97;
   G.~F.~Chapline and N.~S.~Manton,
  Phys.\ Lett.\  {\bf 120B} (1983) 105;
  L.~J.~Romans,
  Phys.\ Lett.\  {\bf 169B} (1986) 374.

\bibitem{gs80842} J.~H.~Schwarz,
  Nucl.\ Phys.\ B {\bf 226} (1983) 269.

\bibitem{agw}
L.~Alvarez-Gaume and E.~Witten,
  Nucl.\ Phys.\ B {\bf 234} (1984) 269.

\bibitem{gs84} M.~B.~Green and J.~H.~Schwarz,
  Phys.\ Lett.\  {\bf 149B} (1984) 117.

\bibitem{bim} C.~Bouchiat, J.~Iliopoulos and P.~Meyer,
  Phys.\ Lett.\  {\bf 38B} (1972) 519.

\bibitem{ghmr} D.~J.~Gross, J.~A.~Harvey, E.~J.~Martinec and R.~Rohm,
  Phys.\ Rev.\ Lett.\  {\bf 54} (1985) 502,
  Nucl.\ Phys.\ B {\bf 256} (1985) 253,
  Nucl.\ Phys.\ B {\bf 267} (1986) 75.

\bibitem{cp} J.~E.~Paton and H.~M.~Chan,
  Nucl.\ Phys.\ B {\bf 10} (1969) 516;
  J.~H.~Schwarz,
  Phys.\ Rept.\  {\bf 89} (1982) 223;
N.~Marcus and A.~Sagnotti,
  Phys.\ Lett.\  {\bf 119B} (1982) 97.

\bibitem{calabiyau} P.~Candelas, G.~T.~Horowitz, A.~Strominger and E.~Witten,
  Nucl.\ Phys.\ B {\bf 258} (1985) 46.

\bibitem{yau} For reviews see:  S.~T.~Yau,
  (AMS/IP studies in advanced mathematics. 9);
  B.~Greene and S.~T.~Yau,
  Providence, USA: AMS (1997) 844 p. (AMS/IP studies in advanced mathematics. 1).

\bibitem{cmap}
S.~Cecotti, S.~Ferrara and L.~Girardello,
  Int.\ J.\ Mod.\ Phys.\ A {\bf 4} (1989) 2475.

\bibitem{strominger}
A.~Strominger,
  Commun.\ Math.\ Phys.\  {\bf 133} (1990) 163.
  For a recent review see:
  S.~Cecotti,
  ``Supersymmetric Field Theories : Geometric Structures and Dualities,''
  Cambridge: Cambridsge Univ. Press (2015).

\bibitem{witten85}
E.~Witten,
  Phys.\ Lett.\  {\bf 155B} (1985) 151.

\bibitem{gkp}
S.~B.~Giddings, S.~Kachru and J.~Polchinski,
  Phys.\ Rev.\ D {\bf 66} (2002) 106006
  [hep-th/0105097].

\bibitem{orbifolds} L.~J.~Dixon, J.~A.~Harvey, C.~Vafa and E.~Witten,
  Nucl.\ Phys.\ B {\bf 261} (1985) 678,
  Nucl.\ Phys.\ B {\bf 274} (1986) 285.

\bibitem{closed_4d}
W.~Lerche, D.~Lust and A.~N.~Schellekens,
  Nucl.\ Phys.\ B {\bf 287} (1987) 477;
  K.~S.~Narain, M.~H.~Sarmadi and C.~Vafa,
  Nucl.\ Phys.\ B {\bf 288} (1987) 551;
I.~Antoniadis, C.~P.~Bachas and C.~Kounnas,
  Nucl.\ Phys.\ B {\bf 289} (1987) 87.

\bibitem{classical_solutions} For a review see: M.~J.~Duff, R.~R.~Khuri and J.~X.~Lu, 
  Phys.\ Rept.\  {\bf 259} (1995) 213
  [hep-th/9412184].

\bibitem{mjd}  M.~J.~Duff,
  Class.\ Quant.\ Grav.\  {\bf 5} (1988) 189.

\bibitem{pkt} E.~Bergshoeff, E.~Sezgin and P.~K.~Townsend,
  Phys.\ Lett.\ B {\bf 189} (1987) 75,
  Annals Phys.\  {\bf 185} (1988) 330.

\bibitem{jp} J.~Polchinski,
  Phys.\ Rev.\ Lett.\  {\bf 75} (1995) 4724
  [hep-th/9510017].
  For reviews see:
  J.~Polchinski, S.~Chaudhuri and C.~V.~Johnson,
  hep-th/9602052;
  J.~Polchinski,
  hep-th/9611050.

\bibitem{fliq}
A.~Font, L.~E.~Ibanez, D.~Lust and F.~Quevedo,
  Phys.\ Lett.\ B {\bf 249} (1990) 35.

\bibitem{witten95} E.~Witten,
  Nucl.\ Phys.\ B {\bf 443} (1995) 85
  [hep-th/9503124].

\bibitem{ht}
C.~M.~Hull and P.~K.~Townsend,
  Nucl.\ Phys.\ B {\bf 438} (1995) 109
  [hep-th/9410167];
  P.~K.~Townsend,
  Phys.\ Lett.\ B {\bf 350} (1995) 184
  [hep-th/9501068].

\bibitem{horwit} P.~Horava and E.~Witten,
  Nucl.\ Phys.\ B {\bf 460} (1996) 506
  [hep-th/9510209],
  Nucl.\ Phys.\ B {\bf 475} (1996) 94
  [hep-th/9603142].

\bibitem{Tduality}  K.~S.~Narain,
  Phys.\ Lett.\  {\bf 169B} (1986) 41;
  K.~S.~Narain, M.~H.~Sarmadi and E.~Witten,
  Nucl.\ Phys.\ B {\bf 279} (1987) 369.
  For a review see:
  A.~Giveon, M.~Porrati and E.~Rabinovici,
  Phys.\ Rept.\  {\bf 244} (1994) 77
  [hep-th/9401139].

\bibitem{orientifolds}
  A.~Sagnotti, in Cargese '87, ``Non-Perturbative Quantum Field
Theory'', eds. G. Mack et al (Pergamon Press, 1988), p. 521,
arXiv:hep-th/0208020;
G.~Pradisi and A.~Sagnotti,
Phys.\ Lett.\ {\bf B 216} (1989) 59;
P.~Horava,
Nucl.\ Phys.\ {\bf B 327} (1989) 461,
Phys.\ Lett.\ {\bf B 231} (1989) 251;
M.~Bianchi and A.~Sagnotti,
Phys.\ Lett.\ {\bf B 247} (1990) 517,
Nucl.\ Phys.\ {\bf B 361} (1991) 519;
M.~Bianchi, G.~Pradisi and A.~Sagnotti,
Nucl.\ Phys.\ {\bf B 376} (1992) 365;
A.~Sagnotti,
 Phys.\ Lett.\  {\bf B 294} (1992) 196
 [arXiv:hep-th/9210127].
 For reviews see: E.~Dudas,
Class.\ Quant.\ Grav.\  {\bf 17} (2000) R41 [arXiv:hep-ph/0006190];
C.~Angelantonj and A.~Sagnotti,
Phys.\ Rept.\  {\bf 371} (2002) 1 [Erratum-ibid.\  {\bf 376} (2003)
339] [arXiv:hep-th/0204089].

\bibitem{double_ft}
W.~Siegel,
  Phys.\ Rev.\ D {\bf 47} (1993) 5453
  [hep-th/9302036];
C.~Hull and B.~Zwiebach,
  JHEP {\bf 0909} (2009) 099
  [arXiv:0904.4664 [hep-th]].

\bibitem{adscft} J.~M.~Maldacena,
  Int.\ J.\ Theor.\ Phys.\  {\bf 38} (1999) 1113
   [Adv.\ Theor.\ Math.\ Phys.\  {\bf 2} (1998) 231]
  [hep-th/9711200];
  E.~Witten,
  Adv.\ Theor.\ Math.\ Phys.\  {\bf 2} (1998) 253
  [hep-th/9802150];
  S.~S.~Gubser, I.~R.~Klebanov and A.~M.~Polyakov,
  Phys.\ Lett.\ B {\bf 428} (1998) 105
  [hep-th/9802109].
  For a review see:  O.~Aharony, S.~S.~Gubser, J.~M.~Maldacena, H.~Ooguri and Y.~Oz,
  Phys.\ Rept.\  {\bf 323} (2000) 183
  [hep-th/9905111].

\bibitem{anticipations}
M.~Flato and C.~Fronsdal,
  Lett.\ Math.\ Phys.\  {\bf 2} (1978) 421;
  E.~Angelopoulos, M.~Flato, C.~Fronsdal and D.~Sternheimer,
  Phys.\ Rev.\ D {\bf 23} (1981) 1278;
  P.~Breitenlohner and D.~Z.~Freedman,
  Annals Phys.\  {\bf 144} (1982) 249.
 M.~Gunaydin and N.~Marcus,
  Class.\ Quant.\ Grav.\  {\bf 2} (1985) L11;
  H.~J.~Kim, L.~J.~Romans and P.~van Nieuwenhuizen,
  Phys.\ Rev.\ D {\bf 32} (1985) 389.

\bibitem{holography} J.~D.~Bekenstein,
  Phys.\ Rev.\ D {\bf 7} (1973) 2333,
  Phys.\ Rev.\ D {\bf 49} (1994) 1912
  [gr-qc/9307035];
  G.~'t Hooft,
  Salamfest 1993:0284-296
  [gr-qc/9310026];
  L.~Susskind,
  J.\ Math.\ Phys.\  {\bf 36} (1995) 6377
  [hep-th/9409089].

\bibitem{attractors}
S.~Ferrara, R.~Kallosh and A.~Strominger,
  Phys.\ Rev.\ D {\bf 52} (1995) R5412;
  [hep-th/9508072].
S.~Ferrara and R.~Kallosh,
  Phys.\ Rev.\ D {\bf 54} (1996) 1514
  [hep-th/9602136];
  S.~Ferrara, G.~W.~Gibbons and R.~Kallosh,
  Nucl.\ Phys.\ B {\bf 500} (1997) 75
  [hep-th/9702103];
  H.~Ooguri, A.~Strominger and C.~Vafa,
  Phys.\ Rev.\ D {\bf 70} (2004) 106007
  [hep-th/0405146].

\bibitem{strom_vafa}
A.~Strominger and C.~Vafa,
  Phys.\ Lett.\ B {\bf 379} (1996) 99
  [hep-th/9601029].

\bibitem{microstates}
For reviews see:
 S.~D.~Mathur,
  Fortsch.\ Phys.\  {\bf 53} (2005) 793
  [hep-th/0502050];
I.~Bena and N.~P.~Warner,
  Lect.\ Notes Phys.\  {\bf 755} (2008) 1
  [hep-th/0701216].

\bibitem{planck}
P.~A.~R.~Ade {\it et al.} [Planck Collaboration],
  Astron.\ Astrophys.\  {\bf 594} (2016) A13
  [arXiv:1502.01589 [astro-ph.CO]],
  Astron.\ Astrophys.\  {\bf 594} (2016) A20
  [arXiv:1502.02114 [astro-ph.CO]].

\bibitem{appl_cosmology}
 R.~Kallosh, A.~Linde and T.~Rube,
  Phys.\ Rev.\ D {\bf 83} (2011) 043507
  [arXiv:1011.5945 [hep-th]];
J.~Ellis, D.~V.~Nanopoulos and K.~A.~Olive,
  Phys.\ Rev.\ Lett.\  {\bf 111} (2013) 111301
   Erratum: [Phys.\ Rev.\ Lett.\  {\bf 111} (2013) no.12,  129902]
  [arXiv:1305.1247 [hep-th]].

\bibitem{cosmo_nl}
I.~Antoniadis, E.~Dudas, S.~Ferrara and A.~Sagnotti,
  Phys.\ Lett.\ B {\bf 733} (2014) 32
  [arXiv:1403.3269 [hep-th]];
S.~Ferrara, R.~Kallosh and A.~Linde,
  JHEP {\bf 1410} (2014) 143
  [arXiv:1408.4096 [hep-th]];
 G.~Dall'Agata and F.~Zwirner,
  JHEP {\bf 1412} (2014) 172
  [arXiv:1411.2605 [hep-th]];
   M.~Galante, R.~Kallosh, A.~Linde and D.~Roest,
  Phys.\ Rev.\ Lett.\  {\bf 114} (2015) no.14,  141302
  [arXiv:1412.3797 [hep-th]];
S.~Ferrara, M.~Porrati and A.~Sagnotti,
  Phys.\ Lett.\ B {\bf 749} (2015) 589
  [arXiv:1508.02939 [hep-th]];
  S.~Ferrara, R.~Kallosh and J.~Thaler,
  Phys.\ Rev.\ D {\bf 93} (2016) no.4,  043516
  [arXiv:1512.00545 [hep-th]];
  J.~J.~M.~Carrasco, R.~Kallosh and A.~Linde,
  Phys.\ Rev.\ D {\bf 93} (2016) no.6,  061301
  [arXiv:1512.00546 [hep-th]].
  For a review see:
  S.~Ferrara, A.~Kehagias and A.~Sagnotti,
  Int.\ J.\ Mod.\ Phys.\ A {\bf 31} (2016) no.25,  1630044
  [arXiv:1605.04791 [hep-th]].

\bibitem{constrained_superfields}
 M.~Rocek,
  Phys.\ Rev.\ Lett.\  {\bf 41} (1978) 451;
  E.~A.~Ivanov and A.~A.~Kapustnikov,
  J.\ Phys.\ A {\bf 11} (1978) 2375;
  U.~Lindstrom and M.~Rocek,
  Phys.\ Rev.\ D {\bf 19} (1979) 2300;
  R.~Casalbuoni, S.~De Curtis, D.~Dominici, F.~Feruglio and R.~Gatto,
  Phys.\ Lett.\ B {\bf 220} (1989) 569;
  Z.~Komargodski and N.~Seiberg,
  JHEP {\bf 0909} (2009) 066
  [arXiv:0907.2441 [hep-th]],
  JHEP {\bf 1007} (2010) 017
  [arXiv:1002.2228 [hep-th]];
S.~M.~Kuzenko and S.~J.~Tyler,
  Phys.\ Lett.\ B {\bf 698} (2011) 319
  [arXiv:1009.3298 [hep-th]].

\bibitem{bsb}
  S.~Sugimoto,
Prog.\ Theor.\ Phys.\  {\bf 102} (1999) 685 [arXiv:hep-th/9905159];
I.~Antoniadis, E.~Dudas and A.~Sagnotti,
Phys.\ Lett.\ {\bf B 464} (1999) 38 [arXiv:hep-th/9908023];
C.~Angelantonj,
Nucl.\ Phys.\ {\bf B 566} (2000) 126 [arXiv:hep-th/9908064];
G.~Aldazabal and A.~M.~Uranga,
JHEP {\bf 9910} (1999) 024 [arXiv:hep-th/9908072];
C.~Angelantonj, I.~Antoniadis, G.~D'Appollonio, E.~Dudas and
A.~Sagnotti,
Nucl.\ Phys.\ {\bf B 572} (2000) 36 [arXiv:hep-th/9911081];
E.~Dudas and J.~Mourad,
  Phys.\ Lett.\ B {\bf 514} (2001) 173
  [hep-th/0012071];
  G.~Pradisi and F.~Riccioni,
  Nucl.\ Phys.\ B {\bf 615} (2001) 33
  [hep-th/0107090].

\bibitem{KKLT}
S.~Kachru, R.~Kallosh, A.~D.~Linde and S.~P.~Trivedi,
  Phys.\ Rev.\ D {\bf 68} (2003) 046005
  [hep-th/0301240];
  S.~Kachru, R.~Kallosh, A.~D.~Linde, J.~M.~Maldacena, L.~P.~McAllister and S.~P.~Trivedi,
  JCAP {\bf 0310} (2003) 013
  [hep-th/0308055];
R.~Kallosh, B.~Vercnocke and T.~Wrase,
  JHEP {\bf 1609} (2016) 063
  [arXiv:1606.09245 [hep-th]].

\bibitem{ss}
J.~Scherk and J.~H.~Schwarz,
  Nucl.\ Phys.\ B {\bf 153} (1979) 61.

\bibitem{ss_closed}
R.~Rohm,
  Nucl.\ Phys.\ B {\bf 237} (1984) 553;
  C.~Kounnas and M.~Porrati,
  Nucl.\ Phys.\ B {\bf 310} (1988) 355;
S.~Ferrara, C.~Kounnas, M.~Porrati and F.~Zwirner,
  Nucl.\ Phys.\ B {\bf 318} (1989) 75;
  C.~Kounnas and B.~Rostand,
  Nucl.\ Phys.\ B {\bf 341} (1990) 641;
  I.~Antoniadis and C.~Kounnas,
  Phys.\ Lett.\ B {\bf 261} (1991) 369;
  E.~Kiritsis and C.~Kounnas,
  Nucl.\ Phys.\ B {\bf 503} (1997) 117
  [hep-th/9703059].

\bibitem{ss_branes}
I.~Antoniadis, E.~Dudas and A.~Sagnotti,
  Nucl.\ Phys.\ B {\bf 544} (1999) 469
  [hep-th/9807011];
  I.~Antoniadis, G.~D'Appollonio, E.~Dudas and A.~Sagnotti,
  Nucl.\ Phys.\ B {\bf 553} (1999) 133
  [hep-th/9812118].

\bibitem{borninfeld}
M.~Born and L.~Infeld,
  Proc.\ Roy.\ Soc.\ Lond.\ A {\bf 144} (1934) 425.

\bibitem{fradkin_tseytlin}
E.~S.~Fradkin and A.~A.~Tseytlin,
  Phys.\ Lett.\  {\bf 163B} (1985) 123;
A.~Abouelsaood, C.~G.~Callan, Jr., C.~R.~Nappi and S.~A.~Yost,
  Nucl.\ Phys.\ B {\bf 280} (1987) 599.

\bibitem{magnetic}
C.~Bachas,
  hep-th/9503030;
  M.~Bianchi and Y.~S.~Stanev,
  Nucl.\ Phys.\ B {\bf 523} (1998) 193
  [hep-th/9711069];
  R.~Blumenhagen, L.~Goerlich, B.~Kors and D.~Lust,
  JHEP {\bf 0010} (2000) 006
  [hep-th/0007024];
  C.~Angelantonj, I.~Antoniadis, E.~Dudas and A.~Sagnotti,
  Phys.\ Lett.\ B {\bf 489} (2000) 223
  [hep-th/0007090];
  M.~Bianchi and E.~Trevigne,
  JHEP {\bf 0508} (2005) 034
  [hep-th/0502147].
  For a review see:
  R.~Blumenhagen, B.~Kors, D.~Lust and S.~Stieberger,
  Phys.\ Rept.\  {\bf 445} (2007) 1
  [hep-th/0610327].

\bibitem{hp}
J.~Hughes and J.~Polchinski,
  Nucl.\ Phys.\ B {\bf 278} (1986) 147;
  J.~Hughes, J.~Liu and J.~Polchinski,
  Phys.\ Lett.\ B {\bf 180} (1986) 370;
 I.~Antoniadis, H.~Partouche and T.~R.~Taylor,
  Phys.\ Lett.\ B {\bf 372} (1996) 83
  [hep-th/9512006].

\bibitem{bi_nlin}
S.~Deser and R.~Puzalowski,
  J.\ Phys.\ A {\bf 13} (1980) 2501;

 S.~Cecotti and S.~Ferrara,
  Phys.\ Lett.\ B {\bf 187} (1987) 335;
  J.~Bagger and A.~Galperin,
  Phys.\ Rev.\ D {\bf 55} (1997) 1091
  [hep-th/9608177];
M.~Rocek and A.~A.~Tseytlin,
  Phys.\ Rev.\ D {\bf 59} (1999) 106001
  [hep-th/9811232];
  S.~M.~Kuzenko and S.~Theisen,
  Fortsch.\ Phys.\  {\bf 49} (2001) 273
  [hep-th/0007231];
  S.~Ferrara, M.~Porrati and A.~Sagnotti,
  JHEP {\bf 1412} (2014) 065
  [arXiv:1411.4954 [hep-th]];
  S.~Ferrara, M.~Porrati, A.~Sagnotti, R.~Stora and A.~Yeranyan,
  Fortsch.\ Phys.\  {\bf 63} (2015) 189
  [arXiv:1412.3337 [hep-th]];
  B.~L.~Cerchiai and M.~Trigiante,
  JHEP {\bf 1610} (2016) 160
  [arXiv:1609.07399 [hep-th]].

\bibitem{gen_geom} For a review see:
M.~Grana,
  Phys.\ Rept.\  {\bf 423} (2006) 91
  [hep-th/0509003].

\bibitem{embedding}
B.~de Wit, H.~Samtleben and M.~Trigiante,
  Nucl.\ Phys.\ B {\bf 655} (2003) 93
  [hep-th/0212239].
See also:
G.~Dall'Agata and G.~Inverso,
  Nucl.\ Phys.\ B {\bf 859} (2012) 70
  [arXiv:1112.3345 [hep-th]],
and references therein.

\bibitem{cerncourier}
S.~Ferrara and A.~Sagnotti, CERN Courier, vol.~57, n.~1, January/February 2017.
\end{thebibliography}
\end{document}